\documentclass[%
 reprint,
superscriptaddress,
nofootinbib,
 amsmath,amssymb,
 aps,
]{revtex4-2}

\usepackage{graphicx}
\usepackage{dcolumn}
\usepackage{bm}
\usepackage{comment}
\usepackage{xcolor}

\begin{document}

\preprint{APS/123-QED}

\title{Magnetic Order and Strain in Hexagonal Manganese Pnictide CaMn$_2$Bi$_2$}

\author{R.H Aguilera-del-Toro} 
\affiliation{%
 Departamento de F\'isica Te\'orica, At\'omica y \'Optica, Universidad de Valladolid, 47011 Valladolid, Spain\\
}%
\affiliation{%
 Donostia International Physics Center (DIPC), 20018 Donostia, Spain\\
}%
\affiliation{%
 Centro de F\'isica de Materiales - Materials Physics Center (CFM-MPC), 20018 Donostia, Spain\\
}%
\author{M. Arruabarrena }\thanks{Corresponding author: mikel.arruabarrena@ehu.eus}
\affiliation{%
 Centro de F\'isica de Materiales - Materials Physics Center (CFM-MPC), 20018 Donostia, Spain\\
}%
\author{A. Leonardo}
\affiliation{%
 Donostia International Physics Center (DIPC), 20018 Donostia, Spain\\
}%
\affiliation{%
 EHU Quantum Center, University of the Basque Country UPV/EHU, 48940 Leioa, Spain\\
}%
\author{Martin Rodriguez-Vega} 
\affiliation{Department of Physics, The University of Texas at Austin, Austin, Texas 78712, USA}

\author{Gregory A.\ Fiete}
\affiliation{Department of Physics, Northeastern University, Boston, Massachusetts 02115, USA}
\affiliation{Department of Physics, Massachusetts Institute of Technology, Cambridge, Massachusetts 02139, USA}
\author{A. Ayuela}\thanks{Corresponding author: a.ayuela@csic.es}
\affiliation{%
 Donostia International Physics Center (DIPC), 20018 Donostia, Spain\\
}%
\affiliation{%
 Centro de F\'isica de Materiales - Materials Physics Center (CFM-MPC), 20018 Donostia, Spain\\
}%

\date{\today}

\begin{abstract}
The manganese pnictide CaMn$_2$Bi$_2$, with Mn atoms arranged in a puckered honeycomb structure, exhibits narrow-gap antiferromagnetism, and it is currently a promising candidate for the study of complex electronic and magnetic phenomena, such as magnetotransport effects and potential spin spirals under high pressure. In this paper, we perform a detailed research of the magnetic properties of CaMn$_2$Bi$_2$ using density functional theory (DFT) combined with the Hubbard U correction and spin-orbit coupling, which accurately describe the magnetic interactions. Our results obtained for a large number of magnetic configurations are accurately captured by a modified Heisenberg model that includes on-site magnetization terms to describe magnetic energy excitations. 
We further investigate the role of the spin-orbit coupling, and find that the magnetic anisotropy of CaMn$_2$Bi$_2$ shows an easy plane, with the preferred magnetization direction being exchanged between axes in the plane by applying small strain values. This strain-tunable magnetization, driven by the interplay between spin-orbit interactions and lattice distortions, highlights the potential for controlling magnetic states in Mn-pnictides for future applications in spintronic and magnetoelectric devices. \end{abstract}

\maketitle

\section{Introduction}

Layered transition metal based pnictides have attracted much interest in materials physics due to their intriguing properties, including magnetism, superconductivity, charge density waves, and remarkably high magnetoresistance \cite{paglione2010high}. 
Notably, the compounds with the presence of Bi in the layers, such as BaMn$_2$Bi$_2$ and BaMnBiF, have emerged as promising counterparts to the Fe-based pnictides \cite{BaMn2Bi2}, offering advantages such as lower band gaps and improved accessibility to metallicity. 
Furthermore, the manganese-based pnictides AMn$_2$Pn$_2$ (A = Ca, Sr, Ba; Pn = P, As, Sb, Bi) compounds exhibit interesting properties related to structural changes, with Ca- and Sr-based materials crystallizing in the CaAl$_2$Si$_2$-type trigonal structure (space group P3m1) and Ba-based compounds adopting the ThCr$_2$Si$_2$-type tetragonal structure \cite{jacobs2023bamn}. 
Among these compound, the manganese pnictide CaMn$_2$Bi$_2$, with a honeycomb structure and a narrow electronic gap, has currently emerged as a promising candidate for the study of complex electronic and magnetic phenomena.

Previous studies identified CaMn$_2$Bi$_2$ as being a hybridization gap semiconductor, where interactions between localized states  $d$ states and a metallic band structure (likely $p$ states) opens a band gap \cite{Gibson}.  
This compound exhibits unusual magnetism and a small band gap, and shows semiconducting transport properties.
The importance of the hybridization of Mn 3$d$ orbitals in influencing the behavior observed in CaMn$_2$Bi$_2$ was highlighted by electronic structure calculations.  
In addition, measurements using electrical and Hall resistivity experiments suggest that CaMn$_2$Bi$_2$ acts as an extrinsic narrow gap semiconductor at low temperatures\cite{Piva}. 
Manganese pnictide was observed to show an activation gap of 20 K, while at high temperatures it exhibits properties of a single-band semimetal.   

Under pressure, it was reported that CaMn$_2$Bi$_2$ undergoes transitions and becomes a metal, which can be due to structural, magnetic and electronic instabilities \cite{Gui}.  
The presence of spin-spirals at high pressures was recently shown, which further increases the magnetic complexity in CaMn$_2$Bi$_2$ \cite{spirals-Marshall}. 
Furthermore, when applying a magnetic field, the compound shows anisotropy and magnetic fluctuation\cite{Kawaguchi}. 
However, a systematic study of the gap, magnetic ordering and strain-tunable anisotropy for CaMn$_2$Bi$_2$ using first principles is still lacking.

In this work, we present a detailed study of bulk manganese pnictide CaMn$_2$Bi$_2$, using density functional theory with Hubbard correction and spin-orbit coupling. 
We then investigate the electronic structure, magnetic ordering, and the magnetic anisotropy under the effects of strain which allows to exchange the magnetization direction. 
We believe that our findings on the unique properties of CaMn$_2$Bi$_2$ provide valuable insights into the mechanisms for the development of spintronics and magnetolectronic devices, and offer a guide for future experimental investigations. 
 
\begin{figure*}[!ht]
\centering
  \includegraphics[height=10cm]{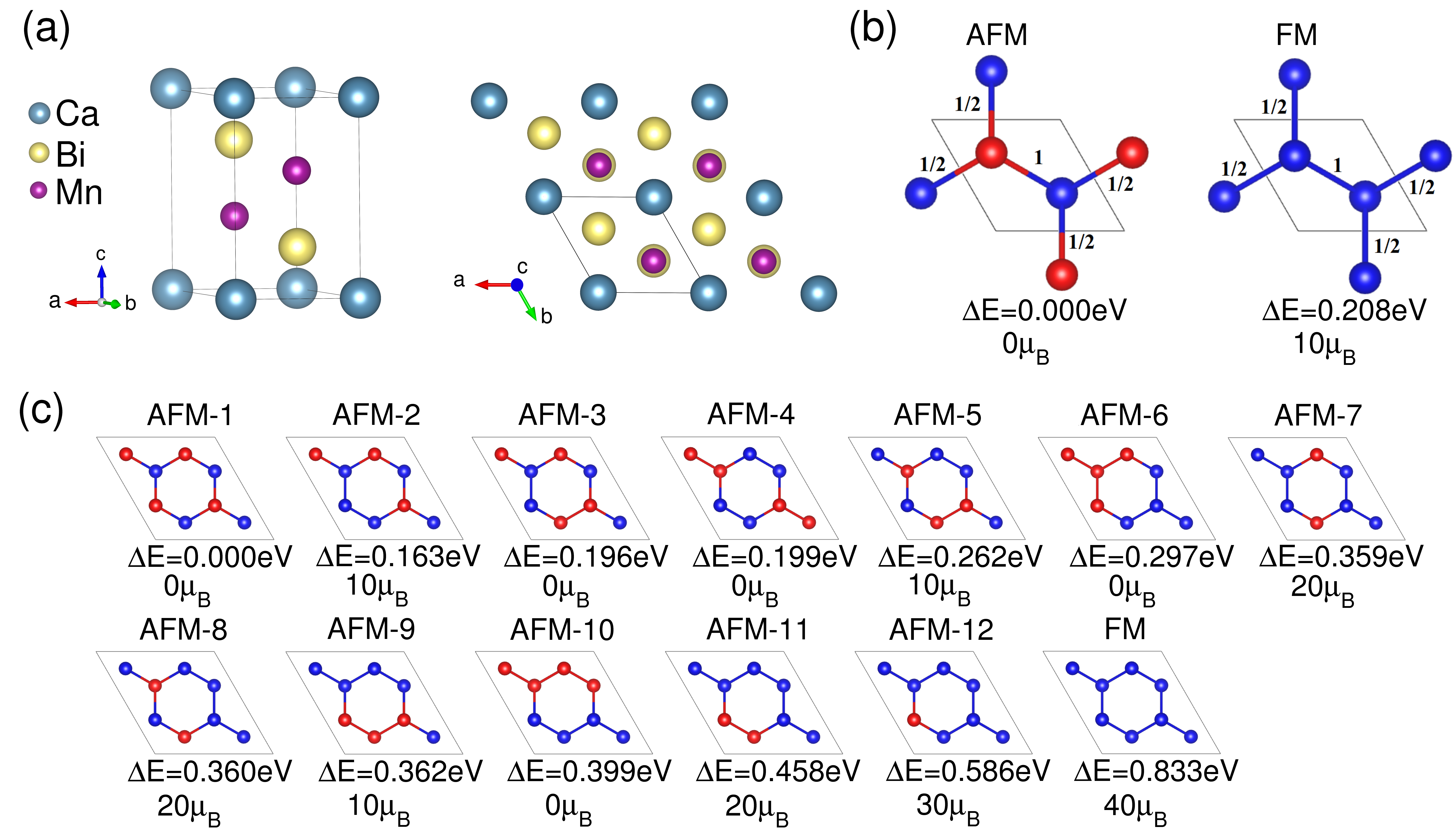}
  \caption{(a) Crystalline honeycomb structure of CaMn$_2$Bi$_2$: (left) side view of the unit cell, and (right) top view of the 2x2x1 supercell. The solid black lines are enclosing the considered unit cell. 
           (b) Scheme of antiferromagnetic and ferromagnetic configurations in the CaMn$_2$Bi$_2$ unit cell. Red (blue) spheres represent Mn atoms with spin up (down). 
           (c) Different magnetic configurations for excitations considering a larger 2x2x1 supercell. The energy differences $\Delta E$ between each configuration the ground state configuration labelled as AFM-1 are given below. The total magnetic moment of each magnetic excitation is also included in $\mu_{\text{B}}$ units.}
  \label{figure1}
\end{figure*}

\begin{figure}[h!]
    \centering
    \includegraphics[scale=0.30]{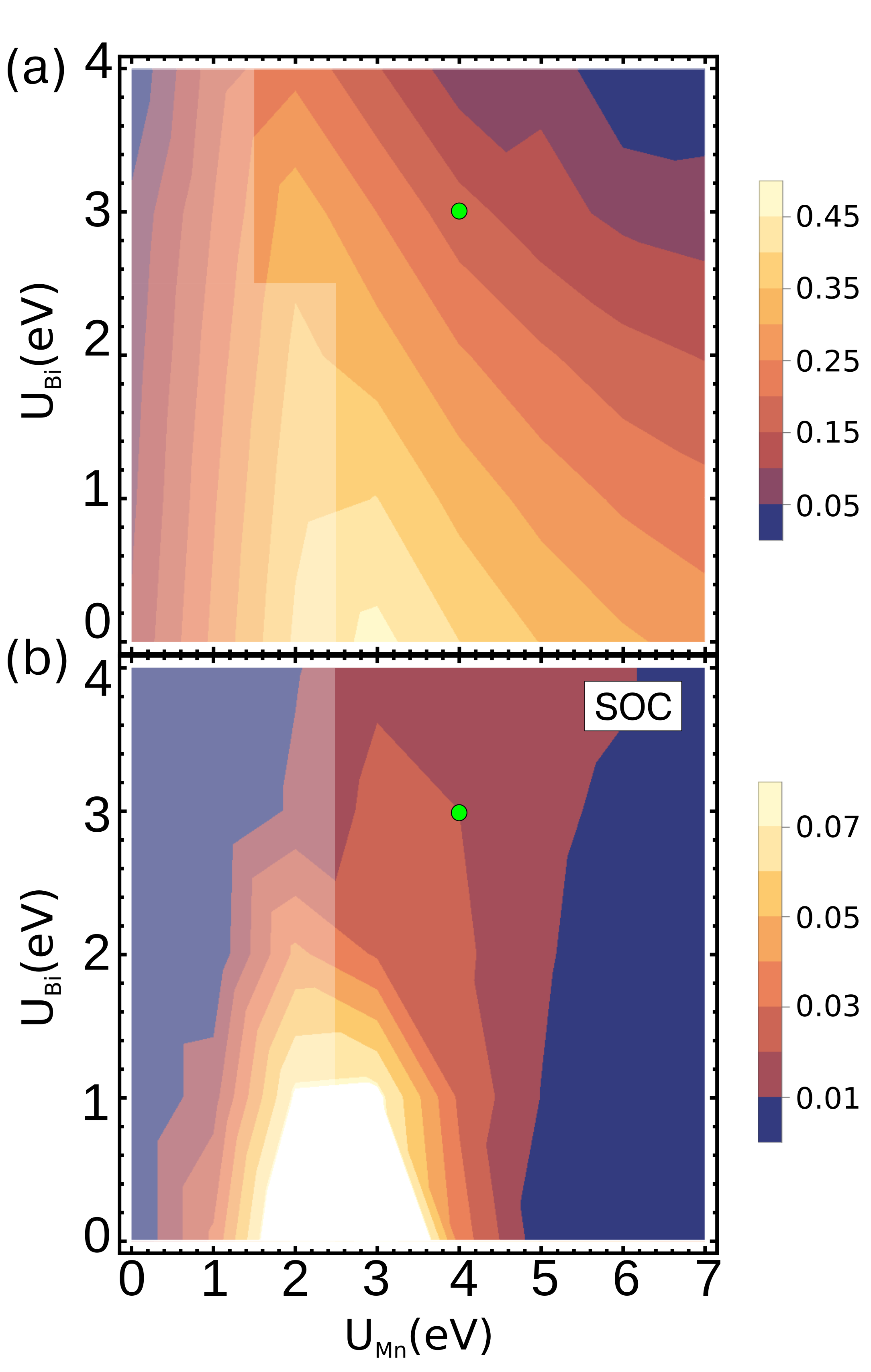}
    \caption{Band gap as a function of the Coulomb parameters $U$$_{\text{Mn}}$ and $U$$_{\text{Bi}}$, (a) without spin-orbit coupling, (b) with spin-orbit coupling. The chosen $U$ values $U$$_{\text{Mn}}$= 4eV and $U$$_{\text{Bi}}$= 3eV used in the calculations in this work are marked with a green circle. 
    The slightly shaded areas depict a region of $U$ values with indirect band gaps between the high symmetry $\Gamma$ and M points. Another region includes the $U$ pairs that show (in)direct band gaps around the $\Gamma$ point (see inset of Fig. \ref{figure-bands} ). 
    }
    \label{gap}
\end{figure}


\section{Theoretical Details}

The side and top views of the CaMn$_2$Bi$_2$ crystal are shown in Fig. \ref{figure1} (a). 
This compound exhibits hexagonal symmetry and is classified in the $P\bar{3}m1$ (164) space group, as determined by experimental measurements \cite{Piva}. 
This manganese pnictide has an antiferromagnetic ground state, with neighbouring Mn atoms in the same layer having opposite spin, as shown schematically in Fig. \ref{figure1} (b). 

The total energy calculations in this paper have been performed in the DFT+$U$ approach, 
using the projector augmented wave (PAW) method implemented in the Vienna Ab-initio Software Package (VASP) \cite{VASP1,VASP2}. 
For the exchange and correlation potential, we use the Perdew-Burke-Ernzerhof form of the generalized gradient approximation (GGA) and incorporated the Coulomb $U$ parameter correction according to the Dudarev GGA+$U$ formulation \cite{Dudarev}. 
To validate our approach, 
we also calculate the band structures of CaMn$_2$Bi$_2$ using the hybrid HSE06 functional as a benchmark. 
We find that the inclusion of the correction terms $U_{\text{Mn}}=4$ eV and $U_{\text{Bi}}=3$ eV yielded band structures in good qualitative agreement with those obtained using the hybrid functional \footnote{However, the hybrid functional greatly overestimates the band gap of the pnictide as discussed below.}. 
 
All calculations were performed with well-converged parameters, including a plane-wave cutoff energy of 700 eV, a gamma-centered 15x15x8 Monkhorst-Pack k-point mesh. 
We relaxed the atomic coordinates until the forces in all directions were less than 0.5 meV/\r{A}. An electronic energy convergence criterion of 10$^{-7}$ eV was applied. 
The atomic valence configurations for Ca, Mn, and Bi were 3$s^2$3$p^6$4$s^2$4$p^{0.01}$, 4$s^2$3$d^5$ and 6$s^2$6$p^3$, respectively.


\section{Results}


\subsection{Electronic properties}

\begin{figure}[h!]
    \centering
    \includegraphics[scale=0.43]{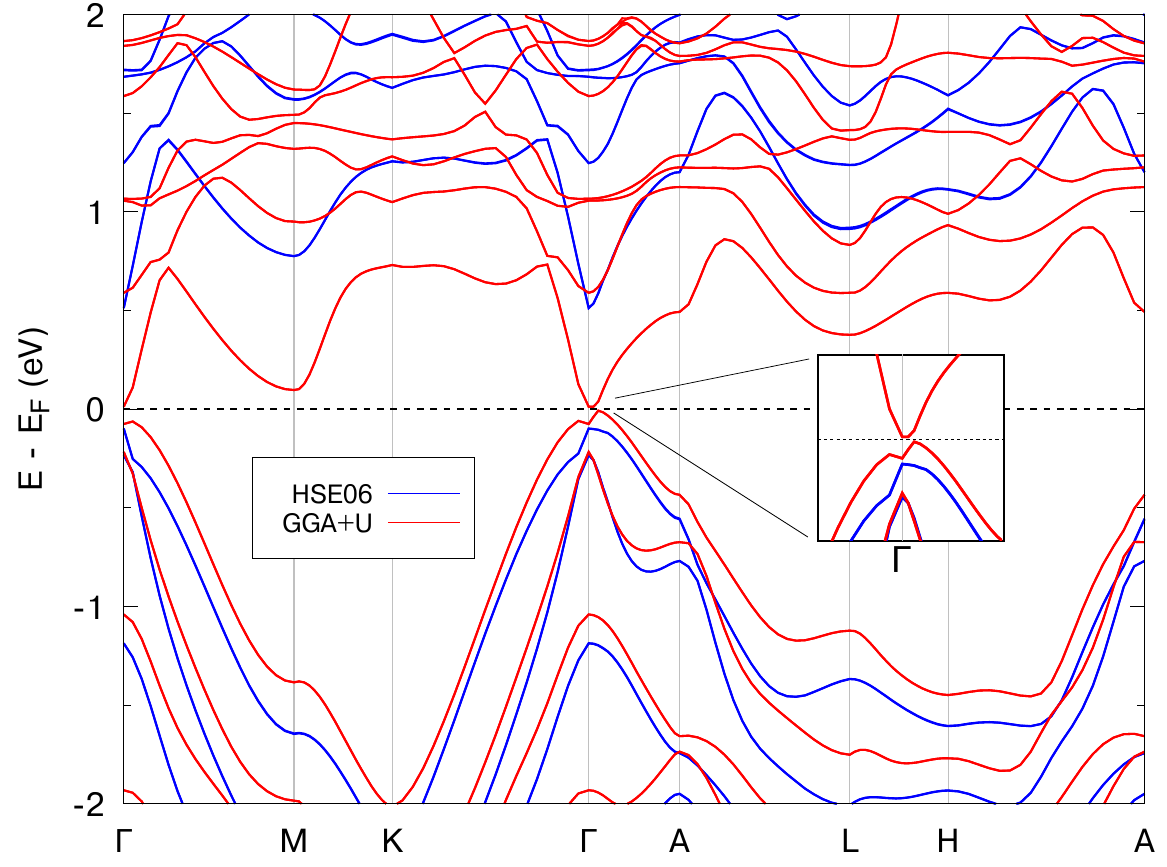}
    \caption{Band structure of CaMn$_2$Bi$_2$ computed using the GGA+$U$ approach (red) and the HSE06 hybrid functional (blue). The inset at the right side shows the indirect band gap around the $\Gamma$ point with more detail.  
    }
    \label{figure-bands}
\end{figure}

We first characterize the electronic properties of manganese pnictide. 
We use several approaches, ranging from the simplest GGA approach to the most computationally demanding hybrid HSE06 calculations including spin$-$orbit coupling (SOC), with a special focus in the GGA+$U$ calculations including SOC. 
We also investigate the effect of the crystalline structure, by comparing the results obtained with the experimental structure and the electronically relaxed structure.

The electronic properties of CaMn$_2$Bi$_2$ are currently under discussion. The experimental band gap values are in the range from 31 to 62 meV, being very small, as determined by transport measurements \cite{Gibson}. 
We find that the band gap 
varies greatly with the numerical method used to calculate it, 
with values ranging from 20 meV (GGA+$U$ with SOC) to 0.7 eV (hybrid HSE06).

In Figure 2, we plot the band gap values as a function of the $U$ parameters: 
panel (a) shows the regular GGA+$U$ calculations, while panel (b) includes the spin-orbit coupling. 
We first highlight the significant effect of the spin-orbit coupling, which narrows the band gap of the compound by splitting the degenerate valence band Bi $p$ and the conduction bands Mn $d$. 
This narrowing can be seen by looking at the scale to the right of each subplot in the figure. We find that the band gaps of manganese pnictide are almost and order of magnitude smaller when SOC is included.
For example, in the GGA+$U$ calculations with $U_{\text{Mn}}=4$ eV and $U_{\text{Bi}}=3$ eV, the SOC reduces the band gap nearly by almost an order of magnitude from 171 meV to 20 meV \footnote{A similar trend is observed in Fig. S3 of Ref. \cite{Lane-2023}}. 
The general trend indicates that the band gap value increases with $U_{\text{Mn}}$ up to a certain point and then decreases. With increasing $U_{\text{Bi}}$ the band gap just decreases. 
It is also important to note that the nature of the band gap depends on the $U$ values. For small values of $U_{\text{Bi}}$, the band gap is indirect, between the $\Gamma$ and M points. For $U_{\text{Bi}}$ values greater than 2.5 eV, it becomes indirect around the $\Gamma$ point.

In Fig. \ref{figure-bands}, the band structure of CaMn$_2$Bi$_2$ is shown, calculated using the GGA+$U$ and hybrid HSE06 approaches. 
The (in)direct band gap around $\Gamma$ can be seen in the inset on the right-hand side of the figure.
We observe that the hybrid calculations are significantly overestimating the band gap, a fact which is in agreement with previous studies\cite{Piva}. However, the band shape is consistent between the hybrid and GGA+$U$ calculations. The bands in the hybrid approach are similar to those including the Hubbard $U$ parameters, which already modify the character of the GGA bands.   
These results suggest that GGA+$U$ calculations including SOC is the preferred method for describing the electronic structure of manganese pnictides, as it provides an accurate description of the bands at a low computational cost. Furthermore, it yields band gap values close to experimental observations with values of $U_{\text{Mn}}=4$ eV and $U_{\text{Bi}}=3$ eV.

Another factor to consider is the role of the crystal structure on the electronic properties of this manganese pnictide. 
We find significant differences in the band gap when using the experimentally or theoretically obtained relaxed structures. 
These differences are due to the structures obtained by DFT+$U$, which tend to be slightly more expanded than the experimental ones. For instance, the lattice parameters for the GGA+$U$ calculation with Hubbard parameters $U_{\text{Mn}}=4$eV and $U_{\text{Bi}}=3$eV are $a=4.77\textup{\r{A}}$ and $c=7.74\textup{\r{A}}$, in contrast to the experimental $a=4.64\textup{\r{A}}$ and $c=7.64\textup{\r{A}}$\footnote{In a recent work, the lattice parameters of $a=4.76\textup{\r{A}}$ and $c=7.72\textup{\r{A}}$ have been obtained for $U=4.75$eV \cite{Lane-2023}.} \cite{Piva}. 
However, we do not find a clear trend between the structure expansion and the compound band gap.


 
\subsection{Magnetic properties}

\subsubsection{Magnetic order}
The magnetic order of CaMn$_2$Bi$_2$ is studied using a 2x2x1 supercell of the chemical unit cell. 
Figure \ref{figure1} (c) shows all the tested magnetic configurations. 
Our findings show that the magnetism arises from the spin density localized around the manganese atoms in the compound. 
The ground state exhibits an antiferromagnetic order between two Mn layers, and has two mutually Mn antiferromagnetic sublattices within each hexagonal layer. 

Since the ground state can be obtained using the chemical unit cell, we proceeded to investigate the Heisenberg spin exchange term. 
Taking into account the crystal symmetry of the system, the exchange coupling strength (J$_e$) can be assumed to remain constant for all Mn-Mn interactions. 
We therefore express the Heisenberg Hamiltonian as follows: 

\begin{equation}
    \small
    H_{Ex}=-\sum_{i,j}^{}J_{e}\textbf{S}_{i}\cdot\textbf{S}_{j}.
    \label{eq:1}
\end{equation}
where $S_i$ and $S_j$ are the localized moments at sites i and j respectively. The total energies of the ferromagnetic (FM) and antiferromagnetic (AFM) configurations in Fig. (\ref{figure1}) are given by  
\begin{equation}
\small
E_{FM}=E_0 + 4(\frac{1}{2}J_{e}S^{2})+J_{e}S^{2}= E_0 + 3J_{e}S^{2}, 
\end{equation}
\begin{equation}
\small
E_{AFM}=E_0 - 4(\frac{1}{2}J_{e}S^{2})-J_{e}S^{2}= E_0 - 3J_{e}S^{2},  
\end{equation}
where $E_0$ represents the total energy of the non-magnetic configuration, and $S=5/2$ corresponds to the spin around each manganese atom.
For the energies, we have considered the interaction between the manganese atoms and their nearest neighbours in the adjacent cells, as shown in Fig. 1c. This gives us the following expression for the exchange coupling (J$_e$):
 
\begin{equation}
\small
J_{e}=\frac{E_{FM}-E_{AFM}}{6S^2} = -5.5 \text{meV},  
\end{equation}
where the negative value indicates the preference for antiferromagnetic couplings between the Mn atoms. 

\begin{figure}[htp]
\centering
  \includegraphics[scale=0.34]{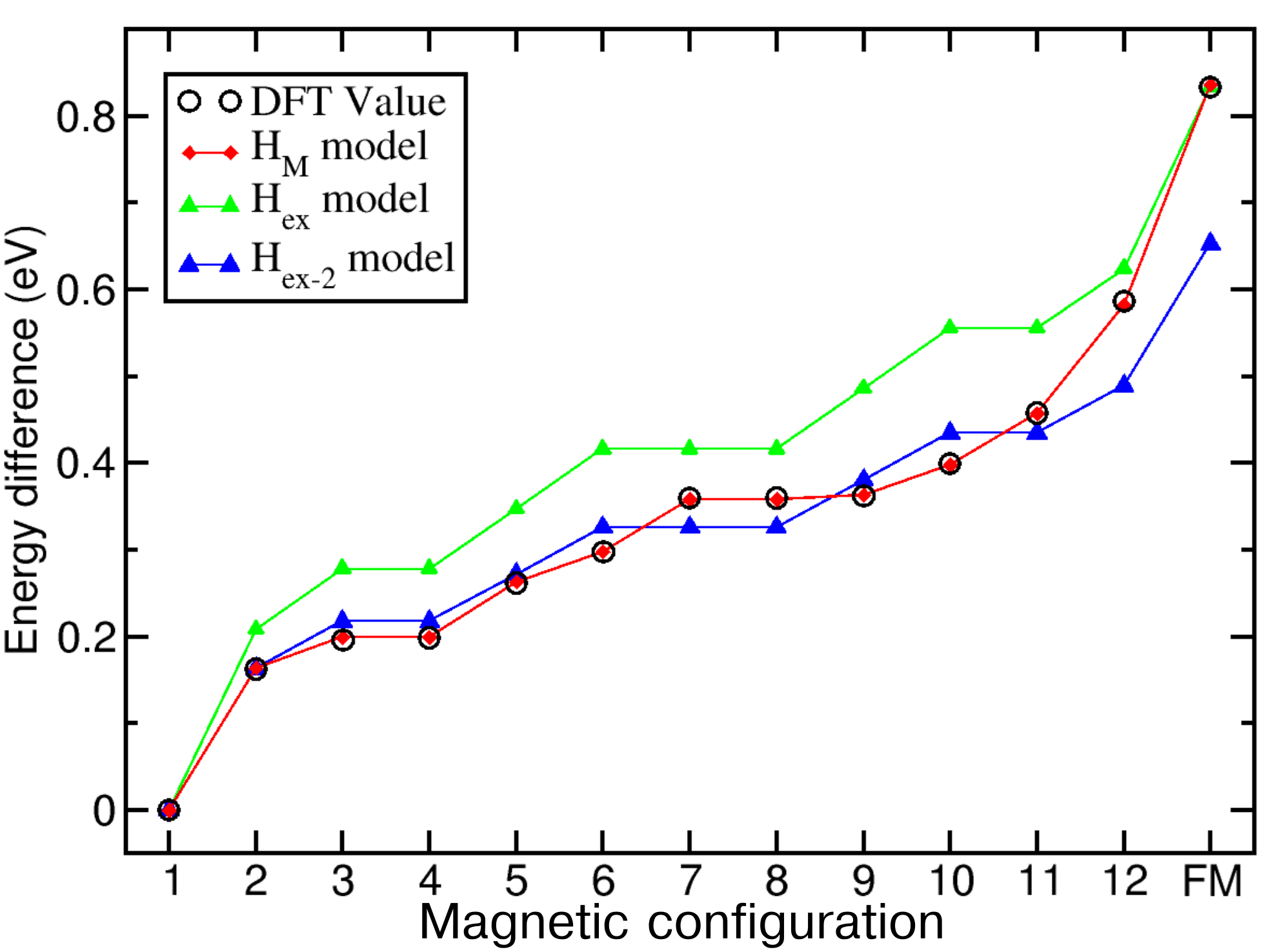}
  \caption{Energy difference between the magnetic excitations in the 2x2x1 supercell and the ground state AFM-1 configuration. 
           The black marks show the energy difference obtained directly by subtracting the DFT total energies. 
            The values in green are calculated using the H$_{\text{ex}}$ model described in Eq. \ref{eq:1}. 
            The energy differences in blue colour are obtained by fitting with the nearest magnetic excitation (H$_{\text{ex-2}}$).  
            The values in red are calculated using the H$_{\text{M}}$ model, which depends on the magnetization of the cell (Eq. \ref{eq:2}).}
  \label{figure4}
\end{figure}

Next the $H_{ex}$ Heisenberg Hamiltonian is used to calculate the energy differences of magnetic configurations in a 2x2x1 supercell, larger than the one previously fitted, as shown in Fig. \ref{figure4}. 
Although the energies of all the other magnetic excitation solutions fall between the AFM and FM solutions, we find significant errors on the order of tenths of an electron volt for most of them. We then investigated other ways os fitting the exchange constant using next nearest neighbours instead of layers. We use the 2x2x1 supercell and fit all the magnetic configuration energies with the nearest magnetic excitation, and then solve for a new Hamiltonian (H$_{Ex-2}$). We obtained an exchange coupling value of 4.350meV. 
The error in the fitted energies remains unacceptably large compared to the original GGA+$U$ results, being of the order of 0.02 eV.  
It seems that relying solely on a Hamiltonian that accounts for the exchange coupling is not enough to provide an accurate description of the magnetic excitations in CaMn$_2$Bi$_2$. 

The inclusion of additional terms, such as the anisotropic spin exchange arising from spin-orbit coupling, would not be able to explain these energy differences, as they are much smaller in energy.
Further corrections to the Heisenberg hamiltonian, such as including the antisymmetric exchange (Dzyaloshinskii-Moriya interaction), would only adjust the energy on the scale in the order of meV. Consequently, the source of the error must be found in other terms.

A closer look at the order of the magnetic excitations in the 2x2x1 supercell allowed us to find an interesting trend. We find that magnetic configurations with different total magnetic moments and the same number of exchanged J$_e$ pairs have similar energy differences in the order of hundredths of meV. 
Thus, this finding suggests that a term related to the total magnetic moment must be included in the Hamiltonian to provide an accurate description of this layered material, as described below. 

\subsubsection{On-site magnetization term}
To address this, we propose the following Hamiltonian:
\begin{equation}     
\small
H_{M}=-J_{M}\sum_{i,j}^{}\textbf{S}_{i}\cdot\textbf{S}_{j}-\frac{M}{N}(\sum_{i}^{}\textbf{S}_{i})^2, 
\label{eq:2}
\end{equation}
where J$_{M}$ and M are constants to be determined and N the number of chemical cells of the system. The first term represents the exchange coupling, while the second term corresponds to the square of the total magnetic moment of the supercell.
Fitting the energy data from the GGA+$U$ calculations to this equation gives the following values for these constants: J$_{M}$ = -3.97meV and M = -2.39meV.

In Figure 4, we compare the energy differences obtained from the GGA+$U$ calculations with the approximation using the Hamiltonians H$_{ex}$, H$_{ex-2}$ and H$_{M}$.
We observe that the curves obtained by considering only the exchange coupling term result in energy values that deviate significantly from the GGA+$U$ calculations. However, when using the Hamiltonian model that incorporates the total magnetic moment of the supercell, we can reproduce these results with an error of tenths of a meV. To validate this model, we have carried calculations on a 3x3x1 supercell \ref{figure5} and compared the obtained energies with those derived from the H$_{M}$ Hamiltonian. In this particular system, the error falls within the range of hundredths to thousandths of an electron volt.

\begin{figure}[htp]
  \includegraphics[height=6.5cm]{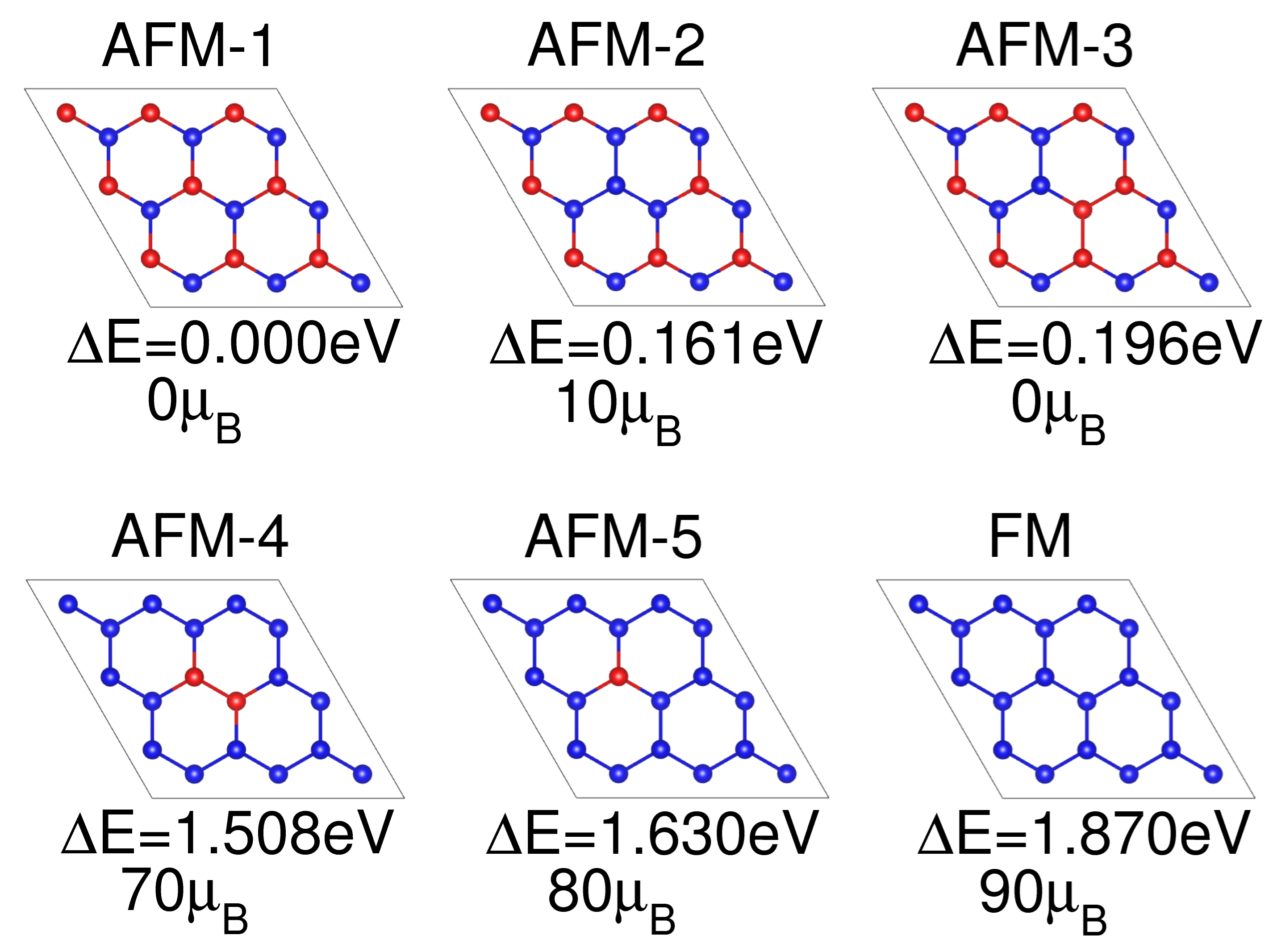}
  \caption{Magnetic excitations for the 3x3x1 cell. Below each case, the energy difference in eV with respect to the ground state and the magnetic moment in $\mu_B$.}
  \label{figure5}
\end{figure}

\subsubsection{Magnetic anisotropy}
Finally, we consider the magnetic anisotropy exhibited by the compound. 
The incorporation of spin-orbit coupling effectively couples the spin magnetic moment to the crystal structure of the compound, leading to magnetic configurations that are either energetically favored or penalized depending on the orientation of the local magnetic moments of the Mn atoms. 
The magnetocrystalline anisotropy energy (MCA) is quantified as the energy difference between a given magnetic direction and the one with the lowest energy. 
We have evaluated the total energy of CaMn$_2$Bi$_2$ for different spin orientations and found that the compound predominantly favors an in-plane configuration, with the Mn spins aligned along the hexagonal a-axis. 
Specifically, the out-of-plane MCA was found to be around 3 meV, while the in-plane anisotropy was around 0.02 meV, highlighting the pronounced easy-plane characteristic of this manganese pnictide. \\

\begin{figure}[]
  \includegraphics[height=6.3cm]{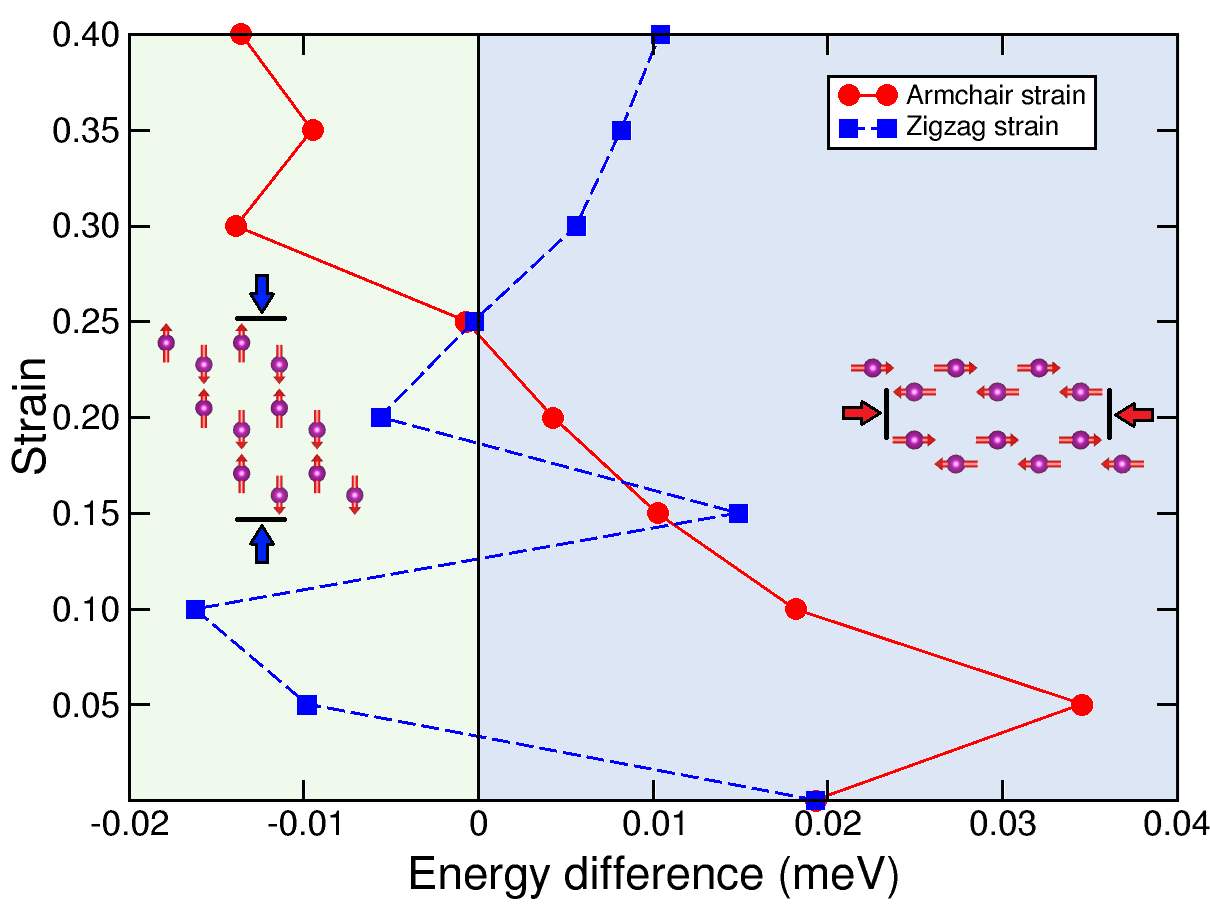}
  \caption{Map of the in-plane magnetic anisotropy energy difference when the strain is applied along the two main hexagonal directions.  The blue region indicates that the zigzag direction is preferred; the green region indicates that the armchair direction is preferred.}
  \label{figure5}
\end{figure}

We also investigated the effect of strain on the magnetic anisotropy. Our analysis shows that the MCA difference with respect to the cartesian $z$-axis (hexagonal $c$-axis) remained constant irrespective of strain, while the disparity in MCA between the in-plane $x$ (hexagonal $a$, armchair) and $y$ (zigzag) directions changed upon the application of strain along the $x$ and $y$-axes, as shown in Fig. \ref{figure5}.  
In particular, for the zigzag direction, the preferred spin orientation of the crystal shifted to align along the $y$-axis at a strain of approximately 0.25\%. 
Subsequently, with further strain applied along the $x$-axis above 0.4\%, the difference in magnetocrystalline anisotropy energy increased, favoring the $y$-axis as the principal easy axis. 
Finally, applying strain in the $y$-direction caused the spin orientation to shift from the zigzag to the armchair direction, with an oscillatory behavior up to a strain of 0.25\%, before returning to the zigzag direction.


\section{Conclusions}

In this work, we have conducted a theoretical investigation of the magnetic properties of the layered pnictide CaMn$_2$Bi$_2$. In the framework of GGA+$U$ including spin-orbit coupling, we have studied the intricate interplay between electronic structure, magnetic ordering, and magnetic anisotropy of the compound. 
Our findings reveal the crucial role of spin-orbit coupling in shaping the electronic band structure and magnetic properties of this material. 
Furthermore, we have proposed a Heisenberg model that accurately captures the magnetic excitations, including those involving multiple unit cells. 
Notably, we have demonstrated that strain engineering can effectively manipulate the magnetic anisotropy, offering a promising path for tuning the magnetic properties of CaMn$_2$Bi$_2$ for potential applications. 
We believe that our results provide valuable insights into the fundamental physics underlying this intriguing material and open the door for future experimental investigations and device development.

\begin{acknowledgments}
We acknowledge funding from the Spanish Ministry of Science and Innovation (grants nos. PID2022-139230NB-I00, and TED2021-132074B-C32), the Gobierno Vasco UPV/EHU (project no. IT-1569-22), the Diputaci\'on Foral de Gipuzkoa (Project No. 2023-CIEN-000077-01), the European Commission  MIRACLE project (GA 964450), and NaturSea-PV (GA 101084348). Research conducted in the scope of the Transnational Common Laboratory (LTC) Aquitaine-Euskadi Network in Green Concrete and Cement-based Materials. 
R.H. Aguilera-del-Toro acknowledges the postdoctoral contract from the Donostia International Physics Center. 
GAF acknowledges support from DOE Award No. DE-SC0022168 and the Alexander von Humboldt Foundation. 
\end{acknowledgments}

\bibliography{CaMn2Bi2.bib}

\end{document}